\documentclass[conference]{IEEEtran}
\IEEEoverridecommandlockouts
\usepackage{cite}
\usepackage{amsmath,amssymb,amsfonts}
\usepackage{algorithmic}
\usepackage{graphicx}
\usepackage{textcomp}
\usepackage{pgfplots}
\usepackage{xcolor}
\usepackage{listings}
\usepackage{verbatim}
\usepackage{tikz}
\usetikzlibrary{shapes, arrows.meta, positioning}
\usepackage{pgf-pie}
\def\BibTeX{{\rm B\kern-.05em{\sc i\kern-.025em b}\kern-.08em
    T\kern-.1667em\lower.7ex\hbox{E}\kern-.125emX}}
\begin{document}

\title{A Framework for Synthetic Audio Conversations Generation 
using Large Language Models\\
}

\author{\IEEEauthorblockN{Kaung Myat Kyaw}
\IEEEauthorblockA{
\textit{Innovative Cognitive Computing (IC2) Research Center} \\
\textit{School of Information Technology} \\
\textit{King Mongkut's University of Technology Thonburi}\\
Bangkok, Thailand \\
kaungmyat.kyaw@kmutt.ac.th}
\and
\IEEEauthorblockN{Jonathan Hoyin Chan}
\IEEEauthorblockA{
\textit{Innovative Cognitive Computing (IC2) Research Center} \\
\IEEEauthorblockA{\textit{School of Information Technology} \\
\textit{King Mongkut's University of Technology Thonburi}\\
Bangkok, Thailand \\
jonathan@sit.kmutt.ac.th}
}
}
\maketitle

\begin{abstract}
In this paper, we introduce ConversaSynth, a framework designed to generate synthetic conversation audio using large language models (LLMs) with multiple persona settings. The framework first creates diverse and coherent text-based dialogues across various topics, which are then converted into audio using text-to-speech (TTS) systems. Our experiments demonstrate that ConversaSynth effectively generates high-quality synthetic audio datasets, which can significantly enhance the training and evaluation of models for audio tagging, audio classification, and multi-speaker speech recognition. The results indicate that the synthetic datasets generated by ConversaSynth exhibit substantial diversity and realism, making them suitable for developing robust, adaptable audio-based AI systems.
\end{abstract}


\section{Introduction} The rise of large language models (LLMs) has transformed natural language processing, enabling coherent, contextually relevant text generation. This progress has improved advancements in various fields, including dialogue systems and automated content creation. However, one under-explored application is generating synthetic audio datasets for multi-speaker conversations, which are essential for training speech-related models. Existing audio conversation datasets are limited in size and diversity, hindering the development of robust models that handle diverse scenarios. To address this, we propose ConversaSynth, a framework for generating synthetic audio conversations using LLMs and text-to-speech (TTS) systems. ConversaSynth produces dialogues with distinct personas covering a wide range of topics, converting them into audio to create a diverse and realistic dataset. This resource supports applications such as:
\begin{itemize} \item \textbf{Audio Classification:} Enhancing the accuracy of models in identifying and categorizing different types of audio content. \item \textbf{Speech Recognition and Extraction:} Improving models’ ability to transcribe and segment multi-speaker conversations. \end{itemize}
ConversaSynth also allows for the creation of specific conversational topics and speaker attributes, filling gaps in existing datasets. This flexibility is vital for training models that perform well across varied contexts and speaker traits. Overall, ConversaSynth addresses current limitations and advances synthetic audio generation for future research and development in machine learning.

\section{Literature Review} The integration of LLMs and TTS technologies has the potential to revolutionize applications such as virtual assistants, content creation, and accessibility tools. This review explores key developments in LLMs and TTS, along with their integration.

\subsection{Large Language Models (LLMs)} LLMs have significantly advanced NLP by leveraging massive datasets to generate human-like text. Progress from GPT-2 to GPT-4 has led to notable improvements in understanding context and generating coherent text \cite{brown2020language}. The Transformer architecture, introduced by Vaswani in 2017 \cite{transformer}, employs self-attention mechanisms to capture long-range dependencies, enabling more sophisticated text generation. OpenAI's GPT-3, with 175 billion parameters, set new benchmarks for LLMs by excelling at various language tasks, and GPT-4 further refined these capabilities. LLMs now power applications like writing assistance, storytelling, and chatbots, demonstrating their versatility in generating high-quality content.
\subsection{Text-to-Speech (TTS) Technologies} TTS technology has evolved from concatenative synthesis, which used pre-recorded speech segments, to deep learning models that generate more natural speech. Tacotron \cite{tacotron} and WaveNet \cite{wavenet} transformed TTS by using mel-spectrograms and neural vocoders, resulting in more realistic synthetic speech. Advanced models like Parler-TTS \cite{parler} and XTTS \cite{xtts} further improved expressiveness and customization, adapting speech output to various contexts. Integrating LLMs with TTS systems, such as using ChatGPT with advanced TTS, has enabled more human-like conversational agents, enhancing personalized virtual assistants and interactive media.

\section{Methodology}
\begin{figure}[h!]
\centering
\begin{tikzpicture}[
    node distance=0.5cm and 0.5cm,
    block/.style={rectangle, draw, text width=12em, text centered, rounded corners, minimum height=3em},
    line/.style={draw, -Stealth}
]

\node (llm_selection) [block] {Selection of Large Language Model (LLM)};
\node (persona_design) [block, below=of llm_selection] {Designing Conversational Personas};
\node (conv_gen) [block, below=of persona_design] {Conversation Generation};
\node (text2speech) [block, below=of conv_gen] {Text-to-Speech Conversion};
\node (audio_concat) [block, below=of text2speech] {Audio Dialogues Concatenation};

\path [line] (llm_selection) -- (persona_design);
\path [line] (persona_design) -- (conv_gen);
\path [line] (conv_gen) -- (text2speech);
\path [line] (text2speech) -- (audio_concat);

\end{tikzpicture}
\caption{ConversaSynth Framework: Methodology Overview}
\label{fig:framework}
\end{figure}

In this section, we outlined the framework, which we coined ConversaSynth, used for generating synthetic audio conversations involving multiple speakers. Our approach is structured around key stages, including the selection of a suitable large language model (LLM), the design of distinct conversational personas, the process of generating conversations, the conversion of text to speech, and the concatenation of audio dialogues as displayed in Fig.1. Each stage is carefully designed to ensure the creation of coherent, contextually relevant, and diverse audio conversations. By leveraging a combination of advanced models and fine-tuned techniques, we aim to produce high-quality synthetic dialogues that maintain consistency in character voices and offer a realistic conversational experience. The following sections detail the methodologies applied at each step, from LLM selection to audio post-processing, to achieve our desired outcomes.

\subsection{Selection of Large Language Model (LLM)}
The selection of an appropriate large language model (LLM) is crucial for generating coherent and contextually relevant conversations. To ensure optimal performance, we conducted an evaluation of several LLMs based on their capabilities and alignment with our requirements. The primary candidates considered were Llama2 \cite{llama2}, Llama3 \cite{llama3}, Gemma2 \cite{gemma2}, and Mixtral \cite{mixtral}. Our selection criteria focused on three key aspects:
\begin{itemize}
    \item \textbf{Performance:} We assessed each model’s ability to generate coherent, contextually relevant, and diverse conversations, ensuring that the generated dialogues are both realistic and engaging.
    \item \textbf{Customization:} We evaluated the flexibility of each model in defining and modifying conversational personas to suit various dialogue styles and topics, which is essential for creating distinct and believable characters.
    \item \textbf{Computation:} We considered the computational resources required to run each model, aiming to balance high performance with efficiency to ensure feasibility and cost-effectiveness.
\end{itemize}

\begin{table}[htbp]
\caption{Time Taken by Different LLMs}
\begin{center}
\begin{tabular}{|c|c|}
\hline
\textbf{LLM} & \textbf{Time Taken (seconds)} \\
\hline
Llama2-7B & 716.84 \\
\hline
Llama3-8B & 530.40 \\
\hline
Gemma2-9B & 579.13 \\
\hline
Mixtral-7B & 3,145.89 \\
\hline
\end{tabular}
\label{tab:time_comparison}
\end{center}
\end{table}

We conducted an experiment to evaluate the performance of various large language models (LLMs) for our work. The experiment involved having the LLMs generate 50 conversation scenarios involving multiple participants, ranging from 2 to 5 individuals. We employed a checksum program to verify that the generated responses adhered to the required format. Following this, we recorded the time taken for response generation and the number of responses that deviated from the correct format. As described in Table \ref{tab:time_comparison}, our findings indicate that Llama3 generates responses more quickly than the other LLMs. Given that our work may require the generation of hundreds or thousands of responses, resource efficiency and the time taken per generation are critical factors for our work. 

\begin{table}[htbp]
\caption{Number of Responses in Incorrect Format by Different LLMs}
\begin{center}
\begin{tabular}{|c|c|}
\hline
\textbf{LLM} & \textbf{Number of Wrong Format Responses} \\
\hline
Llama2-7B & 20 \\
\hline
Llama3-8B & 2 \\
\hline
Gemma2-9B & 0 \\
\hline
Mixtral-7B & 12 \\
\hline
\end{tabular}
\label{tab:wrong_format_responses}
\end{center}
\end{table}

Based on our second measurement in Table \ref{tab:wrong_format_responses}, Gemma2 produced zero responses in the wrong format, while Llama3 had only two incorrect responses. After a comprehensive evaluation, we selected the Llama-3 8B model for conversation generation due to its fast speed and acceptable error rate, ensuring the generation of high-quality and diverse synthetic conversations.

\subsection{Designing Conversational Personas}
To generate diverse conversations, distinct personas are defined with variations in name, characteristics, personality traits, and speaking style. For our sample experiment, we created 9 unique personas to ensure diversity in the generated audio data. Fig.\ref{fig:persona_code} is an example of how a persona is defined in Python.

\begin{figure}[h]
\centering
\begin{lstlisting}
alice = Persona(
    name="Alice",
    characteristics=["Enthusiastic", "Brave", "Curious", "Optimistic"],
    personality="Alice loves exploring unknown territories, meeting new people, and learning about different cultures. Her positive attitude and fearlessness inspire those around her to step out of their comfort zones.",
    style="A woman speaks at a slow pace with very clear audio.")
\end{lstlisting}
\caption{Example of persona definition}
\label{fig:persona_code}
\end{figure}

It is crucial that the persona's speaking style includes the phrase "very clear audio" to ensure the generated audio remains undistorted.

\subsection{Conversation Generation}
First, we randomly determine the number of participants in each conversation, with the number ranging between 2 and 5 individuals. Following this, we randomly select predefined personas corresponding to the chosen number of participants. These selected personas are then added to the beginning of the prompt. To ensure the generated response adheres to the desired format, we utilize few-shot prompting. The resulting prompt is structured as in Fig.\ref{fig:prompt_code}.

\begin{figure}[h]
\centering
\begin{lstlisting}
"<List of selected_personas>
They are all sitting around a table, having a lively and engaging conversation. Always place the whole story inside [CONV_BEGIN] and [CONV_END]. The order of the personas doesn't have to be in sequential order; it could be random. When referring to each character, please put their name in square brackets. Follow the format of the following example: 
Example 1:
[CONV_BEGIN]

[{selected_personas[0].name}] I believe there's a lot to be discussed. 
[{selected_personas[1}.name} ] I agree!

[CONV_END]

Example 2:
[CONV_BEGIN]

[{selected_personas[0}.name}] Sometimes, I think about my life being good.
[{selected_personas[1}.name}] That's great! I envy you.

[CONV_END]"
\end{lstlisting}
\caption{Prompt template for generating conversation text dialogues}
\label{fig:prompt_code}
\end{figure}

The selected\_personas list in Python contains the persona objects that were randomly selected in the previous step. We then perform string manipulation to generate the dialogues for each persona, ensuring that the conversation flows in the correct order as in Fig.\ref{fig:dialogue_example}.
\begin{figure}[h]
\centering
\begin{lstlisting}
{"name": "Cathy", "dialogue": "Did you guys hear about the new comedy club opening up downtown? It's going to be huge!"} 
{"name": "Ben", "dialogue": "\(squinting\) Really? I hadn't heard. What makes you say that?"}
\end{lstlisting}
\caption{Example of dialogues in JSON format}
\label{fig:dialogue_example}
\end{figure}

\begin{figure*}[t]
    \centering
    \includegraphics[width=\textwidth]{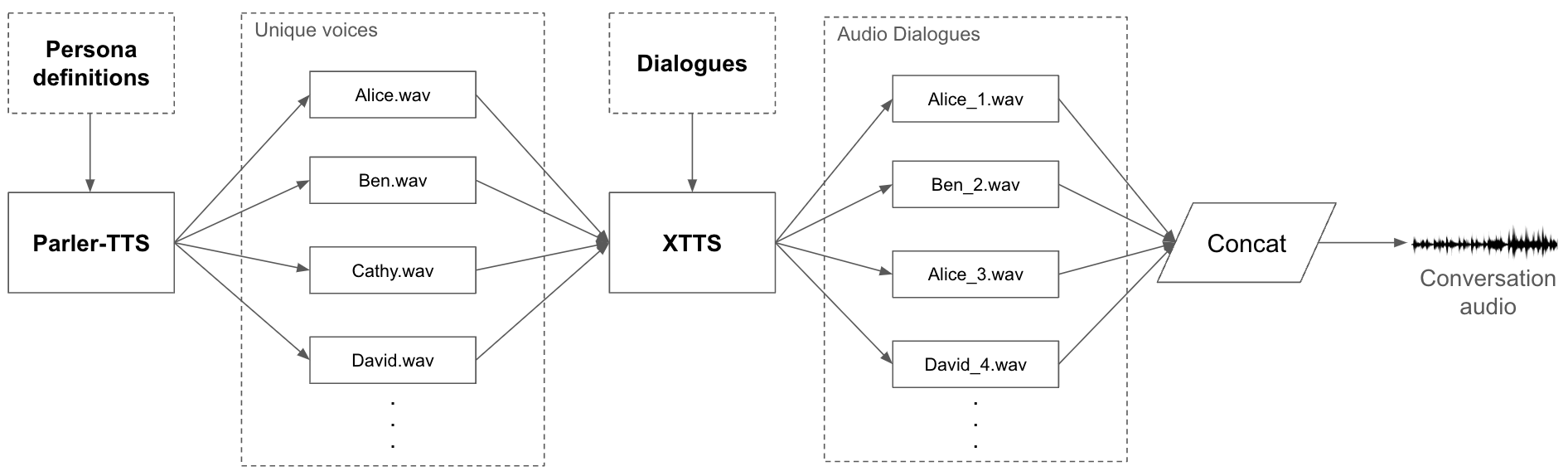}
    \caption{The Flow of Dialogues-to-Audio Conversion in ConversaSynth}
    \label{fig:workflow}
\end{figure*}
As demonstrated, it is still necessary to perform post-processing to address expressions such as \(squinting\) in the example above. These expressions can be problematic, as our text-to-speech model might interpret them as spoken dialogue. To prevent this, we filter out such expressions using regular expressions.

\subsection{Text-to-Speech Conversion}
In this step, the generated text dialogues are converted into audio, with unique voices assigned to each speaker. For text-to-speech conversion, we utilize two different models for distinct purposes. The first model we employ is Parler-TTS, a text-to-speech synthesis model that generates natural-sounding speech from text, delivering high-quality and customizable audio outputs efficiently. Initially, we attempted to use Parler-TTS exclusively for generating the conversation audio. However, Parler-TTS has a limitation: it tends to produce variations in voices even when the same configuration is applied. As a result, the voice of a single persona, such as Alice, may change from one dialogue to another within the same conversation. To address this issue, we incorporate a second model, XTTS, which is a zero-shot text-to-speech model with voice cloning capabilities. Given an audio sample and text input, XTTS can clone the voice from the audio and generate speech using that cloned voice. However, XTTS is limited in its ability to produce multiple unique voices, aside from the few built-in demo voices. To achieve the desired audio quality and consistency, we combine the strengths of both Parler-TTS and XTTS. First, we use Parler-TTS to generate a unique voice for each persona. Next, we employ XTTS to clone these unique voices and generate the speech for each dialogue. The pipeline for generating synthetic conversation audio is illustrated in Fig.\ref{fig:workflow}. This approach ensures that each persona maintains a consistent voice throughout the entire conversation and across the dataset.

\subsection{Audio Dialogues Concatenation}
\begin{table}[htbp]
\caption{Dialogue Segments with Start and End Times}
\begin{center}
\begin{tabular}{|l|c|c|l|}
\hline
\textbf{Filename} & \textbf{Start (s)} & \textbf{End (s)} & \textbf{Speaker} \\
\hline
92.wav & 0.00 & 8.35  & Grace \\
\hline
92.wav & 8.35 & 19.74 & Eva \\
\hline
92.wav & 19.74 & 30.83 & Grace \\
\hline
\end{tabular}
\label{tab:dialogue_segments}
\end{center}
\end{table}

The resulting audio files from the previous text-to-speech conversion step are labeled according to the order of their dialogues, such as Alice\_0.wav, David\_1.wav, and Alice\_3.wav. These dialogues are then concatenated to form a complete conversation audio file, which is saved in .wav format. Following this, a ground truth CSV file is created, which includes the filename, starting timestamp, ending timestamp, and speaker information, as illustrated in Table \ref{tab:dialogue_segments}. Additionally, the conversation audio can undergo further post-processing, such as adding background noise or reverb. 

\section{EXPERIMENT}
In this section, we show the experimental procedures and setups used to evaluate the effectiveness of the synthetic audio conversation generation framework. The experiment was designed to assess the quality and the validity of the generated conversations.

\subsection{Experimental Setup}
We generated 200 synthetic audio conversations with 2 to 5 speakers using Llama3-8B for text generation and Parler-TTS combined with XTTS for voice synthesis and cloning. Nine personas (Alice, Ben, Cathy, Eva, David, Henry, Isabella, Grace, and Frank) were randomly assigned to ensure diversity in dialogue styles and topics. All experiments were conducted on an Nvidia GTX 1080Ti to ensure optimal performance.

\subsection{Challenges and Considerations}
During the experiment, several challenges were encountered. The primary issue was the occasional failure of the Llama3-8B model to generate conversations in the correct format, leading to a small percentage of unusable outputs. 

\section{RESULTS}
The results of our experiment demonstrate the efficacy and robustness of the proposed framework for generating synthetic audio conversations. 

\subsection{Text Generation Time}
During the text generation phase, the Llama3-8B model generated 200 conversation scenarios, involving 2 to 5 participants each. Out of the 200 generated conversations, 189 adhered to the correct format, resulting in a success rate of 94.5\%. The total time taken for this process was 3,730.06 seconds, averaging approximately 18.65 seconds per conversation. 

\subsection{Text-to-Speech Conversion Time}
The process of converting the generated text conversations into audio using Parler-TTS and XTTS took 4,457.94 seconds in total, averaging approximately 22.29 seconds per audio file. The cumulative duration to generate 189 audio conversations was approximately 8,200 seconds in total, and the time required remains within acceptable limits for practical applications.

\subsection{Exploratory Data Analysis} The dataset includes 189 conversations totaling 4.01 hours, suitable for tasks like audio classification and speech recognition. Speaker distribution is fairly balanced, with 2-speaker segments making up 27.5\%, and 3-, 4-, and 5-speaker segments comprising 22.2\%, 25.9\%, and 24.3\%, respectively, as shown in Table \ref{tab:speaker_distribution_pie}.

\begin{table}[htbp]
\caption{Distribution of Speakers per Conversation}
\begin{center}
\begin{tabular}{|c|c|}
\hline
\textbf{Number of Speakers} & \textbf{Percentage (\%)} \\ \hline
2 Speakers                  & 27.5                    \\ \hline
3 Speakers                  & 22.2                    \\ \hline
4 Speakers                  & 25.9                    \\ \hline
5 Speakers                  & 24.3                    \\ \hline
\end{tabular}
\label{tab:speaker_distribution_pie}
\end{center}
\end{table}

In terms of speaker appearances, Table \ref{tab:speaker_distribution} shows "Alice" had the most with 300 appearances, while "Frank" had the fewest at 205.

\begin{table}[htbp]
\caption{Speaker Appearance Distribution}
\begin{center}
\begin{tabular}{|c|c|}
\hline
\textbf{Speaker} & \textbf{Appearance Count} \\ \hline
Alice            & 300                       \\ \hline
Ben              & 278                       \\ \hline
Cathy            & 246                       \\ \hline
Eva              & 238                       \\ \hline
David            & 234                       \\ \hline
Henry            & 229                       \\ \hline
Isabella         & 215                       \\ \hline
Grace            & 206                       \\ \hline
Frank            & 205                       \\ \hline
\end{tabular}
\label{tab:speaker_distribution}
\end{center}
\end{table}

\subsection{Audio Quality Evaluation}

To evaluate the quality of the generated synthetic audio conversations, we calculated the Signal-to-Noise Ratio (SNR) for each audio file. SNR is a commonly used metric in audio quality assessment, measuring the level of the desired signal relative to the noise. A higher SNR indicates better audio quality. In our evaluation using SNR, the average SNR across all evaluated audio files was found to be 93.49 dB, indicating a high level of audio clarity. This result suggests that the synthetic audio generated by our framework maintains excellent fidelity, enhancing its potential for practical applications in various domains.

\section{CONCLUSION}
In conclusion, this work successfully developed and validated a framework for generating synthetic audio conversations using large language models (LLMs) and text-to-speech (TTS) systems. The Llama3-8B model, combined with Parler-TTS and XTTS, provided a robust solution for creating diverse and coherent dialogues with consistent and natural-sounding synthetic voices. The generated sample dataset, comprising over 4 hours of multi-speaker conversation audio, represents a valuable resource for various machine learning and artificial intelligence applications.

This work demonstrates that the integration of advanced LLMs and TTS models can significantly enhance the quality and scalability of synthetic audio data generation. The ability to customize and control conversational personas and dialogue topics allows for the creation of datasets tailored to specific needs, addressing the limitations of existing audio datasets in terms of size, diversity, and contextual richness.
\section{FUTURE WORK}
Based on the positive results of this study, there are several possible directions for future research. Future work could integrate synthetic environmental sounds or background noise to create more realistic conversation scenarios, enhancing the applicability of the generated dialogues for audio classification and speech recognition tasks. Additionally, while this study focused on English, assessing the framework's generalization to other languages is essential. This would involve training LLMs and TTS systems on multilingual datasets to evaluate the quality and coherence of conversations in diverse languages. By addressing these areas, the framework can better meet the growing demands of synthetic audio generation in machine learning and artificial intelligence.

\section{ACKNOWLEDGEMENTS}
We would like to express our sincere gratitude to Nattaya Yongpongsa, Danielle Chan, Ariya Silparcha, and Onwara Silparcha for their invaluable assistance with various tasks throughout this research work. Their support made a significant difference. We also acknowledge the moral support provided by the interns at the Innovative Cognitive Computing Lab (IC2). Additionally, we extend our thanks to the Innovative Cognitive Computing Lab (IC2) for providing the computational resources necessary to conduct the experiments. Their contribution was essential to the success of this work.

\end{document}